# Direct measurement of charge transfer phenomena at ferromagnetic/superconducting oxide interfaces

M. Varela[1,†], A.R. Lupini[1], V. Peña[2], Z. Sefrioui[2], I. Arslan[3,4,§], N.D. Browning[3,4], J. Santamaria[2], S.J. Pennycook[1]

[1]Condensed Matter Sciences Division. Oak Ridge National Laboratory, Oak Ridge, TN 37831-6031

[2]GFMC, Dept. Fisica Aplicada III. Universidad Complutense de Madrid, 28040 Madrid. Spain

[3] Dept. Chemical Engineering and Materials Science. University of California-Davis. One Shields Ave. Davis, CA 95616.

[4]National Center for Electron Microscopy, MS 72-150. Lawrence Berkeley National Laboratory. Berkeley, CA 94720

$YBa_2Cu_3O_{7-x}$/$La_{0.67}Ca_{0.33}MnO_3$ ferromagnetic/superconducting interfaces are analyzed by scanning transmission electron microscopy and electron energy loss spectroscopy with monolayer resolution. We demonstrate that extensive charge transfer occurs between the manganite and the superconductor, in a manner similar to modulation-doped semiconductors, which explains the reduced critical temperatures of heterostructures. This behavior is not seen with insulating $PrBa_2Cu_3O_7$ layers. Furthermore, we confirm directly that holes in the $YBa_2Cu_3O_{7-x}$ are located on the $CuO_2$ planes.

Much attention has been given lately to a new type of ferromagnetic/superconducting F/S oxide heterostructure where a high critical temperature superconductor (HTCS) and a colossal magnetoresistant (CMR) manganite are combined [1-10]. Early experimental studies [6-9] on $YBa_2Cu_3O_{7-x}/La_{0.7}Ca_{0.3}MnO_3$ (YBCO/LCMO) structures suggested a strong F/S interplay resulting of the injection of spin polarized carriers and a proximity effect that may have significant new implications both for basic and applied research. Recent reports have shown that the free carrier response in these superlattices is strongly suppressed as compared to the constituent materials, pointing to a transfer of charge between the ferromagnet and the superconductor [5] which has not been directly measured yet. Due to the coupling of charge, orbital and spin degrees of freedom, charge transfer effects could be quite relevant to the physics of these oxide based F/S heterostructures. Furthermore, since complex oxide interfaces often exhibit nontrivial electronic properties, which in some cases can not be explained in terms of conventional band pictures [11-13], the understanding of charge transfer phenomena needs a careful analysis of the relation between interface structure and electronic properties at the atomic scale. In this letter, we analyze in detail the interface charge transfer in $YBa_2Cu_3O_{7-x}/La_{0.7}Ca_{0.3}MnO_3$ superlattices. We demonstrate significant charge transfer between the ferromagnet and the superconductor. As a result, the value of the hole density in the superconductor at the interface is determined by the number of carriers transferred from the ferromagnet, resulting in a value well below optimal doping within 3 nm of the interface. This effect will dominate the properties of ultrathin YBCO layers in YBCO/LCMO heterostructures.

The samples for this study were high quality YBCO/LCMO and $YBa_2Cu_3O_{7-x}/PrBa_2Cu_3O_7$ (YBCO/PBCO) superlattices grown by high oxygen pressure sputtering [9, 14]. Samples were grown with variable YBCO thickness (1<n<15 unit cells) and a constant LCMO (or PBCO) thickness of 15 unit cells (approximately 5.8 nm). These materials have small lattice mismatch (<0.5%), allowing high quality sample growth. Electron microscopy observations and electron energy loss spectroscopy (EELS) were acquired in a scanning transmission electron microscope (STEM) VG Microscopes HB501UX operated at 100kV and equipped with a Nion aberration corrector and a parallel electron energy loss spectrometer. This microscope is capable of routinely achieving a spatial resolution of 0.1 nm. The image in figure 1(c) was obtained in the VG603 microscope, operated at 300 kV and also aberration corrected, capable of a achieving a spatial resolution below 0.8 nm. Cross sectional samples for STEM were prepared by conventional methods.

The critical temperatures of thin HTCS films sandwiched in between CMR spacers are comparatively lower than those obtained when using a non magnetic material, and heterostructures with ultrathin YBCO layers (one or two unit cells thick) are non superconducting (see figure 1(a)) [9, 14]. Poor structural quality can be ruled out from previous x-ray diffraction [9] and from the present electron microscopy observations. Figure 1(b) shows a low magnification image of a [YBCO$_{3\text{ u.c.}}$/LCMO$_{15\text{ u.c.}}$]$_{100\text{nm}}$ sample, with flat and continuous layers. Figure 1(c) shows a high magnification image of an YBCO/LCMO interface from a [YBCO$_{10\text{ u.c.}}$/LCMO$_{15\text{ u.c.}}$]$_{100\text{nm}}$ sample. For this imaging technique, known as Z-contrast, the intensity of each atomic column is roughly proportional to $Z^2$, giving direct compositional contrast [15]. In YBCO the CuO chain plane is the lightest in the unit cell, and is further away from the BaO plane than the CuO$_2$ layer. It therefore appears significantly darker in the image, while BaO planes show up with the brightest contrast. The YBCO/LCMO interface is coherent and free of defects. Occasionally, some steps one unit cell high are observed, consistent with the small step disorder found by x-rays [9]. A high spatial resolution compositional analysis can be achieved by measuring the changes in the intensity of EELS signals corresponding to the different chemical elements of interest. The changes in the La M$_{4,5}$ edge at 842 eV, Ba M$_{4,5}$ line at 781 eV, Mn L$_{2,3}$ edge at 644 eV, and Ca L$_{2,3}$ edge at 346 eV were monitored by placing the electron beam at the interface and scanning across it (note that these edge energies correspond to nominal values). The intensities of the edges were integrated and normalized to give a quantitative measurement of elemental concentration (Mn, La, Ba and Ca), as shown in figure 2(a). A sharp jump at the interface is observed, which is not atomically sharp because some broadening (5 Å each side of the interface) is expected to be due to beam broadening through the specimen thickness. Figure 2(b) shows the Mn to Cu signal ratio across the interface. Again, the jump is quite sharp (takes place within 3Å). It can be concluded that there is no major chemical interdiffusion, consistent with previous x-ray refinement results [9]. Also, EELS measurements confirm a correct La/Ca ratio of 0.7/0.3 in these samples.

A model accounting for the observed interface structure is shown in figure 1(d). The terminating plane in the manganite is found to be a $MnO_2$ plane and the YBCO growth begins with the BaO plane. Figure 3(a) shows a Z-contrast image of a [$YBCO_{1\ u.c.}$/$LCMO_{15\ u.c.}$]$_{100nm}$ sample. The YBCO layer is marked with an arrow. Its structure, as identified from the images and also by atomic plane by atomic plane EELS, shows an atomic plane stacking of $MnO_2$-BaO-$CuO_2$-Y-$CuO_2$-BaO-$MnO_2$: the CuO chains are missing. It is generally accepted that CuO chains act as the charge reservoirs in YBCO: holes in the superconducting $CuO_2$ planes appear as a result of electron transfer to the CuO chains. Thus, the absence of the CuO chains at the LCMO/YBCO interface plane by itself could explain the lack of superconductivity in such *incomplete* ultrathin YBCO layers.

Surprisingly, spectroscopic measurements with atomic resolution show the existence of a noticeable density of holes within these ultrathin YBCO layers. In HTCS materials, the oxygen 2p bands lie very close to the Fermi energy. The O K edge, which results from exciting transitions from the oxygen 1s core level to the oxygen 2p bands can be used to probe the occupancy of the oxygen 2p bands, i.e. the carrier density in the superconductor [16, 17]. Actually, the pre-peak in the O K edge contains mostly O 2p band contributions. Its intensity can be quantified by fitting the area under the peak to a Gaussian function and normalizing by the area under the main body of the edge, also fitted to a Gaussian [17], allowing a quantification of the density of holes. Figure 3(b) shows the O K edge in an optimally doped sample (top) and at the center of the YBCO layer in the [$YBCO_{1\ u.c.}$/$LCMO_{15\ u.c.}$]$_{100nm}$ (bottom). A noticeable pre-peak was found in the ultrathin YBCO layer, whose intensity is comparable to that of deoxygenated $YBa_2Cu_3O_{6.4}$ [17]. Interestingly, $YBCO_{6.4}$ has a quite low Tc, around 30K, but it is still superconducting. On the contrary, superconductivity is absent in [$YBCO_{1\ u.c.}$/$LCMO_{15\ u.c.}$]$_{100nm}$ superlattices with the same doping level. These EELS measurements were repeated at 100K and the same features were reproduced.

In these interfaces, the terminating $MnO_2$ planes electronically couple the manganite to the YBCO. The white lines at the onsets of the *$L_3$* and *$L_2$* absorption edges of Mn are a characteristic signature of electronic transitions from the *$2p_{1/2}$* and *$2p_{3/2}$* core states to unoccupied d-like states near the Fermi level [18-20]. In particular, the ratio of the intensity of the *$L_3$* peak to the *$L_2$* peak, the $L_{2,3}$ ratio, correlates with the 3d-band occupation, i.e., the formal oxidation state [18-23]. Figure 3(c) shows the Mn valence, as obtained from the $L_{2,3}$ ratio across the LCMO layer. These measurements have been averaged over lateral length scales of the order of 8 nm. The average Mn valence across the layer is 3.37±0.10 (the expected value for the chemical doping in this sample). Interestingly, within the layer the Mn oxidation state is slightly high, around 3.45, while at the interfaces it decreases significantly, to below 3.20. This drop in the value of the Mn oxidation state takes place within 1 nm of the interfaces. In LCMO, the fine structure of the O K edge also gives a measure of the carrier density, given the fact that the first peak in the edge (shown in figure 3(d)) contains quite significant 3d band contribution [24, 25]. This peak is significantly reduced close to the $MnO_2$ interface plane (shown in figure 3(d)), demonstrating a reduced hole density which is consistent with the decrease in the Mn oxidation state. Figure 3(e) shows the prepeak intensity, normalized to the main peak in the edge around 534 eV, as a function of the distance to the interface. Again, we see that the carrier density recovers the bulk value after approximately 1 nm, as observed in the Mn 3d bands occupation. Quantitative EELS measurements show that these interfaces are not oxygen deficient. Therefore, the higher occupation of the Mn 3d bands at the interfaces acts in the right direction to preserve charge neutrality in the vicinity of the incomplete YBCO layer, and suggests the possibility of a band bending scenario due to a Fermi level mismatch in the YBCO and LCMO.

Evidence for significant transfer of charge between LCMO and YBCO was found in YBCO/LCMO superlattices with thicker YBCO layers. Again, it is possible to track the spatial changes in the O K edge by placing the electron beam on the interface and acquiring EEL spectra atomic plane by plane while moving into the YBCO. We will discuss the case of a [YBCO$_{10\ u.c.}$/LCMO$_{15\ u.c.}$]$_{100nm}$ sample. On the interface YBCO unit cell the density of holes takes the value found in the [YBCO$_{1\ u.c.}$/LCMO$_{15\ u.c.}$]$_{100nm}$ sample; approximately 3 nm into the YBCO layer the optimal doping value of bulk YBCO is measured. In between, a smoothly increasing density of holes is found, as shown in figure 4(a). This extended hole filling (and the ensuing depression in the order parameter at the interfaces) indicates a large scale transfer of electrons from the LCMO layers into the YBCO. While the overall qualitative behavior of the Mn valence in the LCMO layers in this sample is like that shown in figure 3(c), *the average Mn formal oxidation state was found to increase with increasing YBCO thickness*. In this sample, its value is +3.5 ± 0.10. If we keep in mind that the nominal chemical doping yields an average Mn valence of +3.3 (the value confirmed by EELS in bulk LCMO samples), around 0.2 electrons per LCMO unit cell are missing. From these EELS measurements, it is possible to spatially map the amount of extra charge within the YBCO layers and the doping level of the LCMO (measured from the Mn formal oxidation state), as shown in figure 4(b). In this figure the amount of extra (depleted) electrons per unit length within the YBCO (LCMO) layers is shown. For hole quantification purposes, the pre-peak intensity in deoxygenated YBCO was used as a calibration [17]. The data used to construct this figure are those shown in figure 4(a) and a data from an independent linescan obtained across the LCMO layer (not shown here). The image on the background is only a guide to the eye. The amount of electrons missing from the LCMO (the area underneath the red curve) is consistent with the amount of extra electrons present in the YBCO by the interface (area underneath the blue curve). Figure 4(b) also seems to point to the presence at some extent of interface charge, which might be related to strains and/or the interface oxygen stoichiometry [26].

The observed lack of uniformity in the Mn oxidation state in these samples, according to the LCMO phase diagram, could reflect an inhomogeneous magnetization in the layers. The decreasing trend measured by EELS in the average Mn 3d band occupation (again, the Mn average valence goes from +3.3 for a YBCO thickness of one unit cell to a value around +3.5 for a YBCO thickness around 10 unit cells) is consistent with an increasing charge transfer from LCMO to YBCO as the YBCO layer thickness increases. Simultaneously, the value of the saturation magnetization in these [YBCO$_{n\ u.c.}$/LCMO$_{15\ u.c.}$]$_{100nm}$ samples decreases when the YBCO layer thickness increases, as shown in figure 4(c) (saturation is reached for YBCO layer thicknesses around 5 unit cells), which could also be an effect of the transfer of electrons into the YBCO interface layers. It is worth noting that, although charge transfer phenomena will modify the superconducting properties over distances within 3 unit cells from the interface (from fig. 4(a)), superconductiviy in YBCO/LCMO superlattices is depressed over a much longer scale compared to samples with non magnetic spacers (YBCO/PBCO superlattices) [9, 14]. This shows that spin injection phenomena are a source of superconductivity depression with a much longer length scale than charge transfer phenomena [27]. On the other hand, the modification of the magnetic properties of LCMO resulting from the electron transfer can have a direct impact on proximity phenomena. The Mn oxidation state of +3.5 renders the LCMO at the vicinity of a critical point where ferromagnetism coexists with antiferromagnetic phases. LCMO layers will be magnetically inhomogeneous, as reflected by the depressed saturation magnetization. The presence of antiferromagnetic regions might provide an explanation for the long range proximity effect [7], which on theoretical grounds is not to be expected in half-metallic ferromagnet/ superconductor junctions.

The situation in the superconducting/insulator (S/I) YBCO/PBCO interface is quite different, as summarized in figure 4(d), where again the hole density in the superconductor vs. the distance to the interface is plotted. In this case, the CuO$_2$ planes in the first YBCO unit cell by the interface already show a doping level close to optimal. No band bending would be expected in these samples due to the isoelectronic environment of the Cu. Interestingly, EELS measurements in these samples show a spatial oscillation of the density of holes with a period equal to the lattice parameter, $c$=11.7Å. The hole density decreases on the CuO chains while it increases on the CuO$_2$ planes to match the bulk value. Corresponding changes were observed in the Cu L edge, showing a modulation of the Cu formal oxidation state: the show a higher oxidation state is found in the CuO$_2$ planes than in the CuO chains. These measurements, which were also reproduced in YBCO thick films, constitute a direct proof that the holes responsible for superconductivity are localized within the $CuO_2$ planes, in agreement with the commonly accepted belief. Also, it is worth noting that the gradient observed in the hole density when going from CuO chains to $CuO_2$ planes within the YBCO unit cell is exactly the same as measured in the YBCO/LCMO interface. Using again the EELS data from deoxygenated YBCO [17] as a reference to quantify hole doping from the O K edge prepeak intensity, we can estimate that the density of holes changes at a rate of 0.17 holes per half YBCO unit cell along the *c*-axis in both the YBCO/LCMO and the YBCO/PBCO cases.

In summary, we have demonstrated significant charge transfer effects at LCMO/ YBCO interfaces through the $MnO_2$ interface plane, as shown by atomic resolution EELS. A 3 nm YBCO layer at the interface shows a significantly reduced hole concentration, while the LCMO layers show a corresponding increase in hole doping. The situation is similar to modulation doped semiconductors except the screening length is much shorter. Since CMR and HTCS oxides are extremely sensitive to doping, these charge transfer processes at the interfaces will directly affect the superconducting and/or magnetic properties of the individual layers. Therefore these factors should be taken into account when trying to understand the interplay between superconductivity and ferromagnetism at oxide F/S interfaces.

**Acknowledgements**

Fruitful discussions with R. Sanchez are acknowledged (MV). This research was sponsored by the Laboratory Directed Research and Development Program of ORNL, managed by UT-Batelle, LLC, for the U.S. Department of Energy under Contract No. DE-AC05-00OR22725 and by appointment to the ORNL postdoctoral Research Program administered jointly by ORNL and ORISE. Funding from U.S. Department of Energy under Contract No. DE-FG02-03ER46057 is also acknowledged.

REFERENCES

[†] Corresponding author: mvarela@ornl.gov

[§] Present address: Department of Materials Science and Metallurgy, University of Cambridge, Pembroke Street, Cambridge, CB2 3QZ, UK.

FIGURE CAPTIONS

Figure 1: (Color online) (a) Dependence of the Tc with YBCO layer thickness for a set of superlattices with a magnetic spacer (LCMO, circles) and a non magnetic spacer (PBCO, squares), both spacers being 60Å thick (Data from reference [9]). (b) Low magnification Z-contrast image of a [YBCO$_{3\ u.c.}$/LCMO$_{15\ u.c.}$]$_{100nm}$ superlattice. (c) High resolution Z-contrast image of a typical YBCO/LCMO interface, obtained in the VG603 (from reference [28]). The dotted line points out the interface position. (d) Structural model to account for the interface structure observed in fig. 1(c). Note how the interface CuO chains of YBCO are missing (marked with an arrow).

Figure 2: (Color online) (a) Normalized elemental concentration of La (triangles), Ca (diamonds), Mn (circles) and Ba (squares) in a YBCO(right)/LCMO(left) interface as a function of the distance to the interface (Data on La, Ba and Mn from reference [10]). (b) Mn/Cu ratio across the interface. In both figures the vertical dotted line marks the interface position. The sketches on top of both figures represent the interface structure on a matching scale, according to the color code of figure 1(d).

Figure 3: (Color online) (a) High resolution Z-contrast image of a [YBCO$_{1\ u.c.}$/LCMO$_{15\ u.c.}$]$_{100nm}$ superlattice. The YBCO layer has been marked with an arrow. One YBCO unit cell has been marked with a rectangle. (b) EELS showing the oxygen K edge as obtained from bulk YBCO (top) and the center of the YBCO layer in the [YBCO$_{1\ u.c.}$/LCMO$_{15\ u.c.}$]$_{100nm}$ superlattice (bottom). The bulk spectrum has been displaced vertically for clarity. The pre-peak position has been highlighted with a dotted line. (c) Mn formal oxidation state across the LCMO layer for the same sample. The data have been averaged over a lateral length of 8 nm. Dotted vertical lines mark the interface locations. (d) Oxygen K edge taken at the center of the LCMO layer (bottom) and on the interface plane (top). The pre-peak has been marked with a dotted line. The interface spectrum has been displaced vertically. (e) Normalized pre-peak intensity as a function of the distance to the interface within the LCMO layer.

Figure 4: (Color online) (a) Normalized pre-peak intensity for the O K edge acquired at the YBCO side of a YBCO/LCMO interface, as a function to the distance to the interface. The horizontal dotted line marks the pre-peak intensity in bulk YBCO (Data from reference 28). (b) Amount of excess (depleted) electrons per formula within the YBCO (LCMO) layers as a function of distance along the growth direction, as measured from EELS. Red and blue lines are a guide to the eye. The vertical black lines represent the interface positions. (c) Saturation magnetization measured at 5K for a set of [YBCO$_{n\ u.c.}$/LCMO$_{15\ u.c.}$]$_{100nm}$ superlattices vs. YBCO thickness. The dotted line is a guide to the eye. (d) Normalized pre-peak intensity within the YBCO in a YBCO/PBCO interface, as a function to the distance to the interface (Data from reference [29]). All of the background images are in real scale in all cases, they are only meant to be a guide to the eye. They were not acquired simultaneously with the EELS spectra.

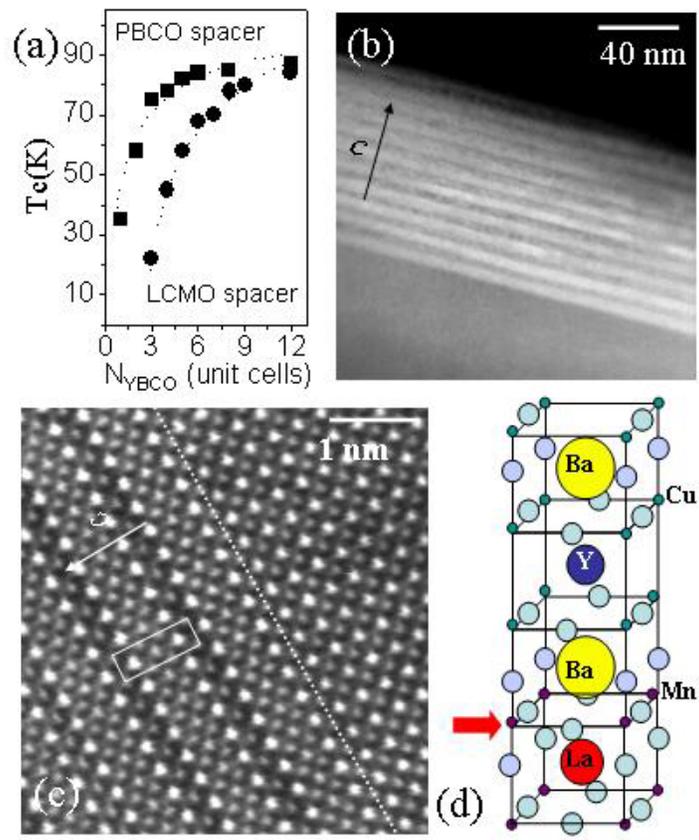

Figure 1

M. Varela *et al.*

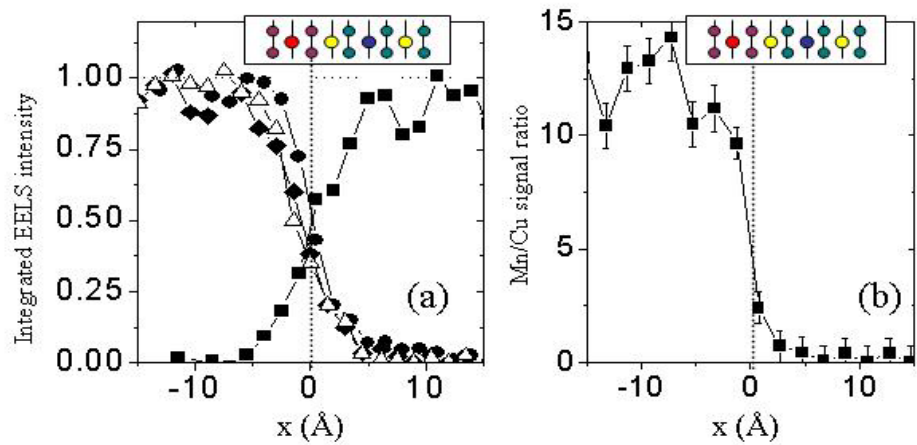

Figure 2

M. Varela *et al*

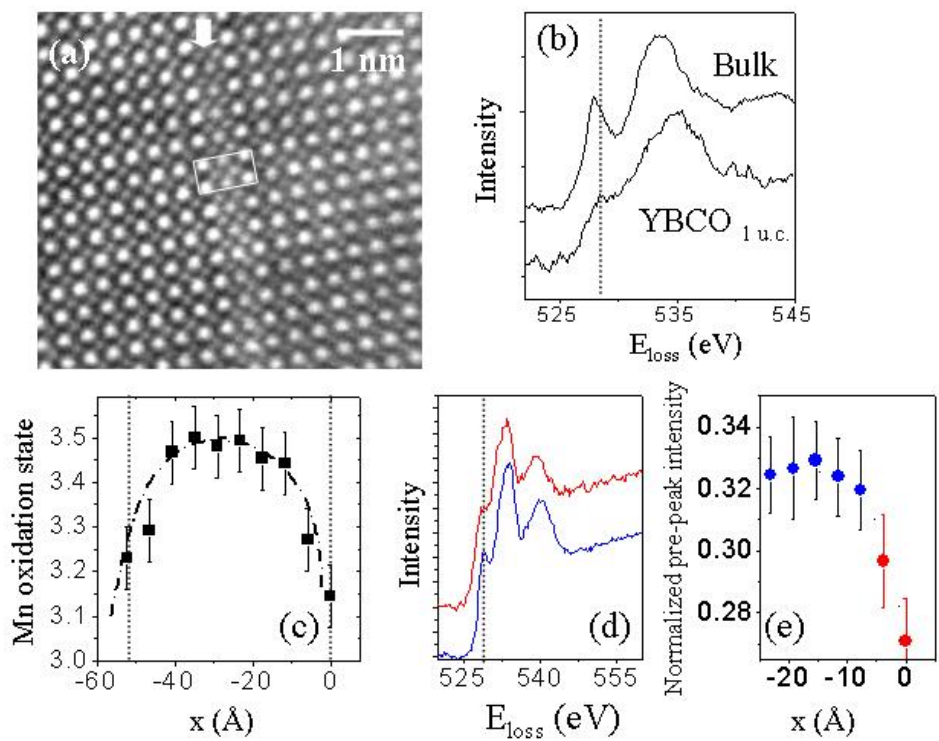

Figure 3

M. Varela *et al.*

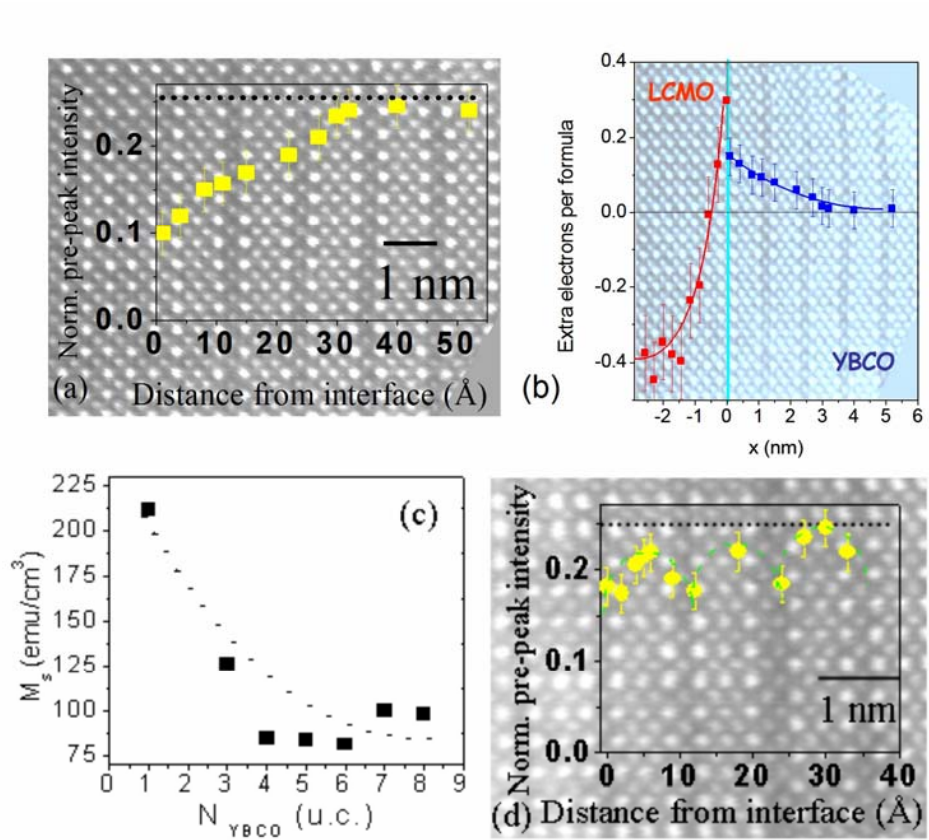

Figure 4

M. Varela *et al.*